\documentclass[useAMS,usenatbib]{mn2e}
\usepackage{graphicx}                                            
\usepackage{times}
\usepackage{epsfig}
\usepackage{rotating}

\newcommand{\he}[1] {He\,{\sc #1}}
\newcommand{\hel}[2] {He\,{\sc #1}~$\lambda$#2}

\newcommand{\hb}{H$\beta$}
\newcommand{\hg}{H$\gamma$}

\newcommand{\ionl}[3] {#1\,{\sc #2}~$\lambda$#3}
\newcommand{\ion}[2] {#1\,{\sc #2}}
\def\kms{\mbox{${\rm km}\:{\rm s}^{-1}\:$}}

\def\lesssim{\mathrel{\hbox{\rlap{\hbox{\lower4pt\hbox{$\sim$}}}\hbox{$<$}}}}
\let\la=\lesssim
\def\gtrsim{\mathrel{\hbox{\rlap{\hbox{\lower4pt\hbox{$\sim$}}}\hbox{$>$}}}}
\let\ga=\gtrsim

\def\sol{~\mathrm{M}_\odot}

\def\kem{K_\mathrm{em}}
\title{Dynamical constraints on the neutron star mass in EXO 0748-676}
\author[T. Mu\~noz-Darias et al.]{T.~Mu\~noz-Darias$^1$ \thanks{E-mail: tmd@iac.es}, J.~Casares$^1$, K.~O'Brien$^2$, D.~Steeghs$^{3,4}$, I.~G.~Mart\'\i{}nez-Pais$^{1,5}$,\newauthor R.~Cornelisse$^1$, and P.~A.~Charles$^6$\\
$^1$ Instituto de Astrof\'isica de Canarias, 38200 La Laguna, Tenerife, Spain\\
$^2$ European Southern Observatory, Casilla 19001, Santiago 19, Chile\\
$^3$ Dept. of Physics, Univ. of Warwick, Coventry CV4 7AL, UK\\
$^4$ Harvard-Smithsonian Center for Astrophysics, Cambridge, MA 02138, USA\\
$^5$ Departamento de Astrof\'\i{}sica, Universidad de La Laguna, E-38206 La Laguna, Tenerife, Spain\\
$^6$ South Africa Astronomical Observatory, P.O. Box 9. Observatory 7935, South Africa\\}
\begin{document}
\maketitle
\begin{abstract}

We present VLT intermediate resolution spectroscopy of UY Vol, the optical
counterpart of the LMXB X-ray burster EXO 0748-676. By using Doppler tomography 
we detect narrow components within the broad \he{ii} $\lambda4542$, $\lambda4686$ 
and $\lambda5412$ emission lines. The phase, velocity and narrowness of these 
lines are consistent with their arising from the irradiated hemisphere of the donor 
star, as has been observed in a number of LMXBs. Under this assumption we  provide the first dynamical constraints on the 
stellar masses in this system. In particular, we measure 
$K_\mathrm{2}>K_\mathrm{em}=300\pm10$~\kms. Using this value we derive $1\sol \leq M_{1} \leq 2.4\sol$ 
and $0.11\leq q \leq 0.28$. We find $M_{1}\geq1.5\sol$ for the case of a 
main sequence companion star. Our results are consistent with the presence of a massive neutron star as has been suggested by \citet{Ozel2006}, although we cannot discard the canonical value of $\sim 1.4\sol$.
\end{abstract}
\section{Introduction}
EXO 0748-676 (=UY Vol) is a Low Mass X-ray Binary (LMXB) discovered as a 
transient source by EXOSAT in 1985 (\citealt{Parmar1985}). Instead of returning to the quiescent state the system had remained in an active state ever since, showing typical properties of persistent LMXBs. However, it has returned to the off-state very recently \citep{Hynes2008}.\par 
EXO 0748-676 also exhibits irregular X-ray dips and periodic eclipses which 
have provided an excellent determination of the orbital period (P=3.82 h) and the orbital 
inclination ($i=75^{\circ}-82^{\circ}$; \citealt{Parmar1986}). On the other hand, 
\citet{Villarreal2004} have detected a burst oscillation at 45 Hz during type I 
X-ray burst events which is associated with the neutron star (NS) spin frequency. 
Recently, \citet{Ozel2006} have suggested the presence of a massive NS with 
$M_{1}=2.10\pm0.28\sol$ and $R_{1}=13.8\pm1.8$ km in EXO 0748-676. This result is based on the gravitational redshift measured in O and 
Fe absorption lines arising from the surface of the NS (\citealt{Cottam2002}) 
and the assumption that the strong thermonuclear X-ray bursts displayed by the 
system (\citealt{Gottwald1986}) can be modelled by a phase of symmetric 
expansion of the NS radius and a cooling phase where the X-ray emission is Eddington-limited. In contrast, \citet{Pearson2006} have reported \he{ii} $\lambda4686$ Doppler tomograms which show a spot consistent
with the expected gas stream position. These authors favor a NS of $1.35\sol$ 
by fitting different gas stream trajectories to the position of the spot.\par
In this paper we present VLT-FORS1 spectroscopy of the $V\sim17$ optical 
counterpart (UY Vol) of EXO 0748-676 which reveals the presence of narrow emission 
components within the \he{ii} emission lines. In recent years, narrow emission 
lines from the donor have been detected in several LMXBs (see 
\citealt{Cornelisse2008} for details, also \citealt{Casares2009}) since they 
were first discovered by \citet{Steeghs2002} in Sco X-1. 
The Roche lobe filling donor star intercepts the energetic photons from the 
inner accretion disc resulting in the observed optical emission lines from its 
surface. The Doppler motion of these lines traces the orbit of the companion star, thereby 
providing the first constraints on the binary parameters for persistent systems where the photospheric light from the companion is otherwise swamped by the bright accretion disc. It is remarkable that this 
technique has proved the presence of a black hole with a mass $>6\sol$ in 
GX339-4 (\citealt{Hynes2003}; \citealt{md08}), presented evidence for a massive 
NS in Aql X-1 (\citealt{Cornelisse2007}) and X1822-371 (\citealt{Casares2003}; 
\citealt{md05}), and provided the first contraints on the binary parameters in 
six other LMXBs, including the prototypical system Sco X-1 
(\citealt{Steeghs2002}). In this letter we present the first dynamical 
constraints on the stellar masses in EXO 0748-676 based on the study of the narrow 
components present within the \he{ii} $\lambda$4541, 4686 and 5412 emission 
profiles. 
In section 2 we describe the observations and data reduction process, whereas the data analysis is presented in section 3. Results are discussed in sections 4 and 5. 

\section{Observations and data reduction}
We observed EXO 0748-676 on 22 Jan 2008 using the \textit{FOcal Reducer and low 
dispersion Spectrograph-1} (FORS1) attached to the VLT-UT2 (Kueyen) telescope 
at the Observatorio Astron\'omico Cerro Paranal (Chile). We used the 
GRIS\_1200g holographic grating obtaining a total of 28 spectra of 600s each. The 
slit width was fixed at 0.51-arsec resulting in a spectral resolution of $\sim 
90$ km s$^{-1}$ and a wavelength coverage of $\lambda\lambda4185-5600$~\AA. The 
seeing stayed between 0.4-0.6-arsec during the first 4 hours but it increased 
up to 1-arsec during the last hour of the run, making slit losses significant 
and absolute flux calibration unreliable for this part of the night. In total 
we covered $\sim 1.3$ binary orbits (i. e. $\sim 5$ h of data).
The images were de-biased and flat-fielded and the spectra extracted using 
conventional {\sc iraf} routines. The wavelength calibration was performed by 
fitting a fourth-order polynomial to an arc image taken during daytime. This 
resulted in a dispersion of 0.73 \AA~ pix$^{-1}$ and rms scatter $\leq0.04$ 
\AA. We used the sky spectra to correct for velocity drifts due to 
instrumental flexure which were found to be $\leq 10$ km s$^{-1}$ (i. e. 
$\leq 0.25$ pix). The zero point of the wavelength scale was established from the 
position of the strong OI~$\lambda5577.338$ skyline. The spectra were finally 
flux calibrated using observations of the flux standard LTT 1788 and exported 
into our analysis software ({\sc molly}).
 \begin{figure}
\includegraphics[width=8cm]{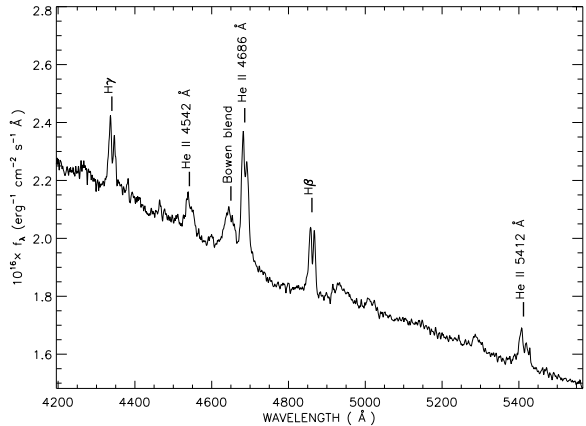}
\caption{Average spectrum of EXO 0748-676 after combining  21 exposures obtained with 
good seeing conditions ($\lesssim 0.6$ arc-sec) and out of eclipse. The most 
prominent lines are indicated for clarity.
\label{a22}}
\end{figure}  
\begin{figure}
\begin{center}
 \includegraphics[width=8cm]{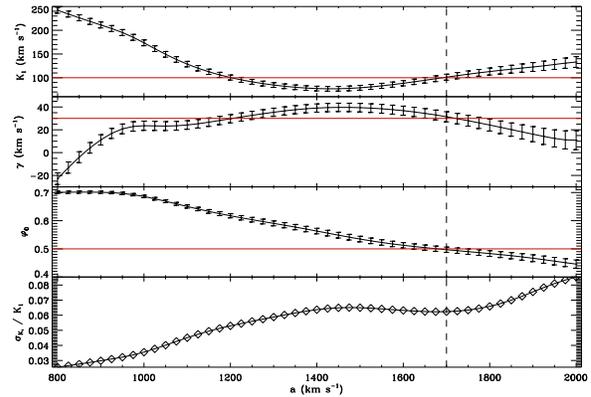}
  \caption{Diagnostic diagram of \hel{ii}{4686} obtained by fitting eq. \ref{dde} to the radial velocity curves 
extracted for each value of the Gaussian separation $a$. The vertical dashed line shows the selected value for $a$ and the horizontal lines the values adopted for each parameter.}
\label{dd}
\end{center}
\end{figure}
\begin{figure}
\begin{center}
  \includegraphics[height=8cm,angle=-90]{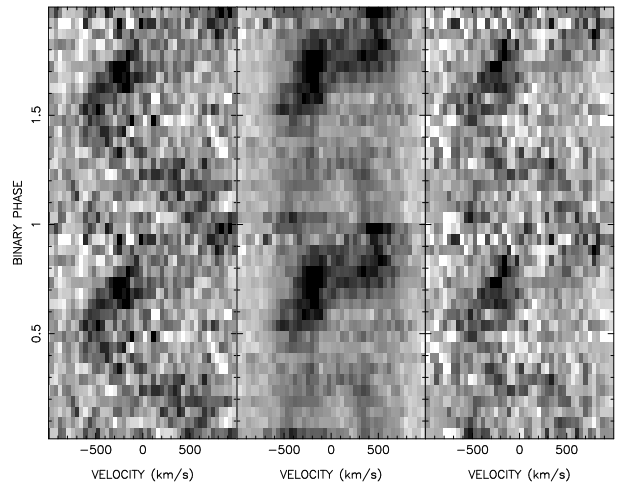}
  \caption{Trailed spectra showing the orbital evolution of the emission lines 
  \he{ii} $\lambda4542$, $\lambda4686$ and $\lambda5412$ (from left to right) 
  in 20 phase bins. Two orbital periods have been plotted for clarity.}
\label{trails}
\end{center}
\end{figure}
\begin{figure*}
\begin{center}
  \includegraphics[height=16cm,width=5.5cm,angle=-90]{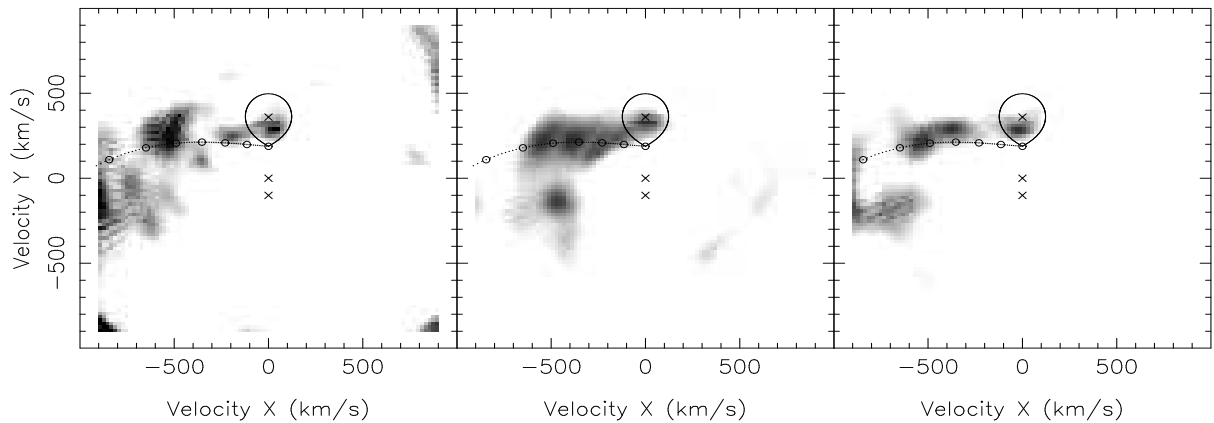}
  \caption{Doppler maps corresponding to \he{ii} $\lambda4542$, $\lambda4686$, 
  $\lambda5412~$ (from left to right) computed using our refined ephemeris. A 
  systemic velocity of $\gamma=70$~\kms was used for the three cases. 
  We have over-plotted the position of the compact object, the gas stream 
  trajectory and the Roche lobe of the companion star assuming $K_1=100$~\kms 
  and $K_2=360$~\kms.}
\label{mdop}
\end{center}
\end{figure*}
\section{Data Analysis}
Fig. \ref{a22} shows the average  spectrum of EXO 0748-676. In order to produce this 
figure we used the 21 out-of-eclipse spectra taken with good seeing conditions, 
comparable to the slit-width. The \hg~and \hb~double peak emission lines 
together with the strong \he{ii} $\lambda4542$, $\lambda4686$, $\lambda5412$ 
features and the Bowen blend at $\lambda\lambda4630-50$ are clearly seen in 
the spectrum. \he{i} $\lambda4922$, $\lambda5016$, and \ionl{Fe}{ii}{5284} are 
also detected, although they are less prominent. In this paper we will focus 
on the analysis of the \he{ii} emission lines where narrow components are found. 
In order to compute the orbital phase of each spectrum we used the 
constant-period ephemeris given by \citet{Wolff2002}. The zero-phase has been 
refined by fitting the mid-eclipse time of an eclipse observed by RXTE on 30 
Jan 2008 (i.e. only 8 days after our observations). We find an error of 
$\sim 0.012$ in orbital phase which has been corrected in the ephemeris. We 
finally adopted:\\
\begin{center}
$T_0(HJD)=2446111.57285(3) +0.1593378191(1)\times N$
\end{center}
 \subsection{Systemic velocity} \label{sdd}
As in our previous work, we decided to apply the double gaussian technique 
\citep{Schneider1980} to the wings of the \hel{ii}{4686} emission line in order to 
obtain a first estimate of the orbital parameters of the NS. We have used 
relative Gaussian separations between $a=800-2000$~\kms in steps of 25~\kms. 
This range has been chosen to avoid low velocity components associated with the 
companion star and the outer disc. After fitting the expression
\begin{equation}
\label{dde} 
V(\phi)=\gamma+K_1\sin{2\pi(\phi-\phi_0)}
\end{equation}
to the radial velocity curves 
extracted for each value of $a$, we obtain the diagnostic diagram presented in 
Fig. \ref{dd}. We adopt the value of $a\sim 1700$~\kms where the phase is correct and also happend to occur near the plateau in the diagnostic parameter $\sigma(K_1)/K_1$ \citep{Shafter1986}. This results in 
$\gamma=30\pm10$~\kms and $K_1=100\pm20$~\kms.\par

This method assumes that the emission line gas orbits the compact object and correctly traces its orbital dynamics. Non-isotropic emisison and/or significant disk asymmetries are known to lead to significant distortions in the radial velocity curves and thus the inferred $K_1$. For this reason the values provided by the diagnostic-diagram 
should be taken as a rough approximation to the   real ones, as will be discussed in the next section.

\subsection{Doppler mapping}  
Extracting radial velocities of faint short-period systems is not 
straightforward. The situation is even more difficult when dealing with narrow 
spectral lines which require high spectral resolution. The spectral resolution 
also degrades with exposure time due to orbital smearing, whereas the 
signal-to-noise ratio (S/N) obviously increases for longer integrations. 
Therefore, in order to avoid significant orbital smearing one has to use short 
exposures (600 s in our case) which results in data sets with good orbital 
sampling but poor $S/N$.\par

Using Doppler tomography \citep{Marsh1988} instead of standard radial velocity 
curves has the benefit of using all the spectra at the same time, allowing one 
to map the brightness distribution of a weak emission line in velocity space. 
This method has been used in almost all previous studies of narrow emission 
lines (see e.g. \citealt{Casares2006}), and only for Sco X-1 (with $V\sim12$ 
and P=18.9 h) has the application of both techniques separately (i.e. radial velocity curves and Doppler mapping) been possible.  In that case a good agreement between the two methods was found \citep{Steeghs2002}.\par

Fig. \ref{trails} shows the evolution of the \he{ii} $\lambda4542$, 
$\lambda4686$ and $\lambda5412$ spectral lines along the orbit. An S-wave with 
a maximum radial velocity of $\sim 500$\kms at orbital phase $\sim 0.5$ 
appears in the 3 maps. 
This translates into bright spots at 
$V(V_\mathrm{X},V_\mathrm{Y})$=(-500,200)~\kms in the Doppler maps of the three 
\he{ii} lines (Fig. \ref{mdop}). Similar spots were also found by \citet{Pearson2006} in several emission lines (e.g  \he{ii} $\lambda4686$). However, in this study we also detect in the three maps compact spots at the position of 
the donor (i.e. $V_X=0$ \kms). The velocities associated with these spots are in the range $V_Y=270-339$ \kms and 
are shown in table \ref{kems}.\par  
We have computed Doppler maps for a wide range of $\gamma$ values bracketing 
$\gamma=30$~\kms, the value favored by the diagnostic diagram. We swept 
between $\gamma=-50$ and 100~\kms  in steps of 10~\kms. 
The effect of using a wrong value of $\gamma$ when computing the maps is to 
obtain out-of-focus spots, which become compact when approaching the correct 
value. By fitting 2D gaussians to the spots as a function of $\gamma$, we favor 
systemic velocities in the range $20\leq \gamma \leq 80$ \kms. We finally 
selected $\gamma=70$~\kms because it produces significantly more compact spots in 
the three maps. In any case, the velocity measurements (table \ref{kems}) 
derived with $\gamma=70$~\kms are consistent within the errors with those 
obtained with $\gamma=30$~\kms.\par

The $\kem$ velocity of a given irradiation line depends on the disc thickness 
and how transparent it is for the irradiating photons. The three \he{ii} lines 
that we study here have similar ionization potentials and, hence, one 
should expect them to show the same $\kem$ value. This is consistent within 
the error bars with the results that we have obtained. Therefore, we decided to
take the weighted mean value of the three lines as our best estimate of $\kem$, 
which results in $K_\mathrm{em}=310\pm10$~\kms. Hence, we will use 
$\kem=300~$\kms as 1-sigma lower limit to $K_2$. 
\section{Reprocessed emission from the companion}
\begin{figure}
\begin{center}
  \includegraphics[width=8cm]{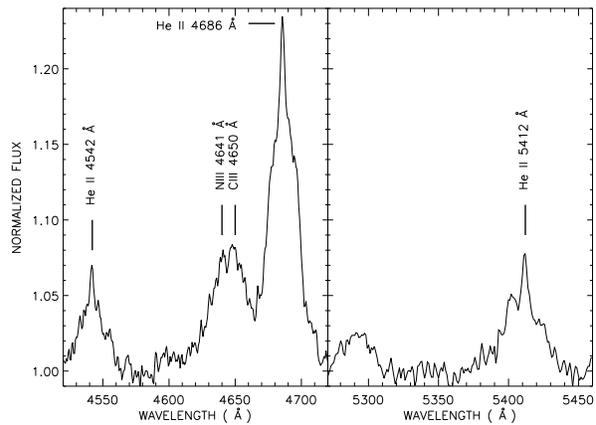}
  \caption{Normalized average spectrum of EXO 0748-676 in the rest frame of the 
  companion (using $K_\mathrm{em} = 300$ \kms). Note the narrow components 
  standing out in \he{ii} $\lambda4542$, $\lambda4686$ and $\lambda5412$.}
\label{rfc}
\end{center}
\end{figure}

Using our $\kem$ value derived from the \he{ii} lines and the refined orbital 
ephemeris it is possible to remove the orbital velocity from each spectrum and compute the average spectrum in the rest frame of the companion. 
Fig \ref{rfc} shows this spectrum, where narrow components standing out in the 
broad \he{ii} $\lambda4542$, $\lambda4686$ and $\lambda5412$ emission lines are 
clearly seen. As an independent test to the Doppler analysis, we have computed 
the spectrum in the rest frame of the donor sweeping a wide range of $\kem$ 
values. We find that, for a range of $\kem$ values around 
$\sim310\pm10$~\kms, the narrow components show   show the highest intensities together with narrow widths consistent with the spectral resolution of $\sim 90$ km s$^{-1}$. 
\begin{table}
\begin{center}
\caption{$K_\mathrm{em}$ $(V_\mathrm{X},~V_\mathrm{Y})$ measurements (\kms) in EXO 0748-676.}
\label{kems}
\begin{tabular}{c c c}
\hline
$\lambda4542$ & \he{ii} $\lambda4686$ & \he{ii} $\lambda5412$ \\
$(28, 300) \pm 18$& $(7, 325) \pm 14$& $(-14, 290)\pm 20$\\
\hline
\end{tabular}
\end{center}
\end{table}
Taking into account the position of the spots in the Doppler maps (i.e. 
orbital phase and velocity), the narrowness of the components 
and the presence of similar high excitation features in all the persistent 
LMXBs that we have studied (see \citealt{Cornelisse2008}) we associate these 
features with X-ray reprocessing from the irradiated hemisphere of the 
donor.\par 

It is also noteworthy that we have a tentative detection of the \ion{N}{iii} and 
\ion{C}{iii} components within the Bowen blend in the rest frame of the 
companion star (left panel in Fig. \ref{rfc}). The weakness of these lines is 
probably due to the reduced number of spectra. We only cover one orbit and at
least two orbits of data are usually required to clearly detect these
transitions (e.g. see \citealt{Casares2006}). However, the \he{ii} emission 
lines are very strong in this system compared to others previously studied and show narrow components that 
provide us with the opportunity to measure $\kem$. It is especially interesting 
to note that EXO 0748-676 is the only system in the Bowen survey where 
\hel{ii}{4542} is as strong as the Bowen blend. \hel{ii}{4686} emission from 
the donor was also reported by \citet{Steeghs2002} in Sco X-1, with a $\kem$ 
value similar to that obtained from the Bowen lines. However, in other 
bright systems where high signal-to-noise studies have been performed (e.g. 
X1822-371 \citealt{Casares2003}) no \he{ii} emission from the donor was 
found. 
\section{System parameters}
EXO 0748-676 offers important advantages with respect to other persistent LMXBs. 
The binary inclination is a known function of the mass ratio owing to the X-ray eclipses \citep{Hynes2006}. 
For instance, $i=82^{\circ}$ yields $q=0.12$, whereas $i=75^{\circ}$ implies $q=0.4$. 
Moreover, since X-ray eclipses and X-ray bursts from the NS surface are 
observed, the NS must be visible, and hence the disc opening angle $\alpha$, 
which obscures part of the irradiated face of the donor, is also related to 
the inclination through $\alpha \leq 90^{\circ}-i~.$  
Therefore for a given $q$, the inclination angle and an upper limit to 
$\alpha$ are set. This is very useful in order to compute the K-correction.\\
Emission lines formed on the inner hemisphere of the donor star as a result of 
X-ray reprocessing only provide a lower limit ($K_{em}$) to the true 
$K_{2}$-velocity. In \citet{md05} we model the deviation between the 
light-center of the reprocessed lines and the center of mass of a Roche lobe 
filling star in a persistent LMXB, including screening effects by a flared 
accretion disc (i.e. the K-correction). We find that this correction depends 
on $q$, $\alpha$ and weakly on the orbital inclination. We now apply the 
K-correction to EXO 0748-676 as a function of $q$, including the above limits on 
$q$ and $i$.
\subsection{The mass of the neutron star}
\begin{figure}
\begin{center}
  \includegraphics[width=8cm]{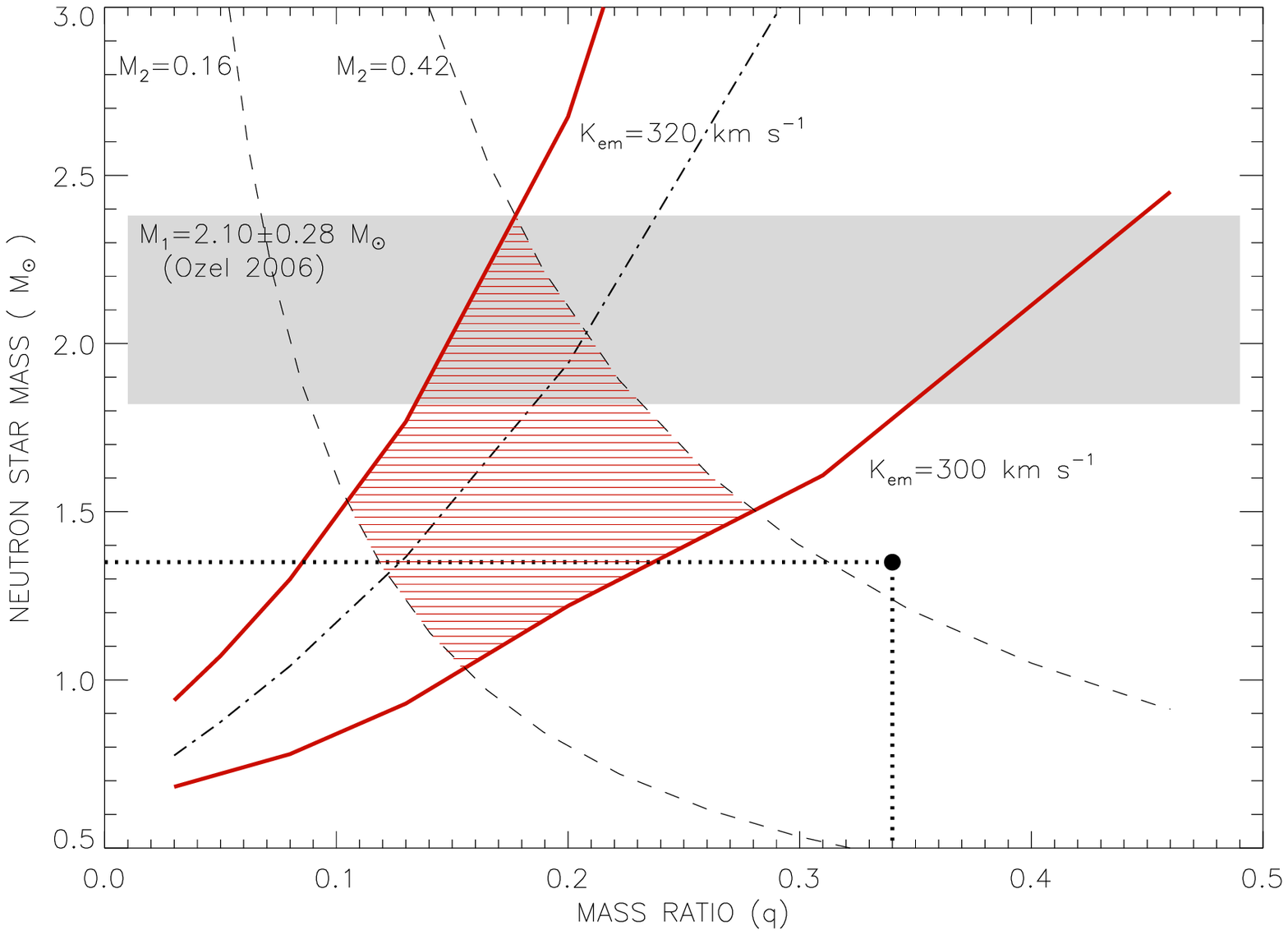}
  \caption{$M_{1}$ vs $q$ obtained from $K_\mathrm{em}=300\pm10$~\kms. The 
  solid lines represent the lower and upper limits to the NS mass for
  $\alpha_max$ and $\alpha=0^{\circ}$ respectively, whereas the dotted-dashed 
  line corresponds to the upper limit for $\alpha \geq 6^{\circ}$ \citep{Meyer1982}. 
  The grey region shows the mass range determined by \citet{Ozel2006} and the 
  solid circle with dotted lines the values favored by \citet{Pearson2006}. 
  The dashed lines are obtained assuming that the donor is either a 
  $M_2\sim0.42\sol$ main sequence star or it is considerably evolved before 
  the onset of mass transfer. The shadowed region indicates our constraints on $q$ and $M_1$.}
\label{masa}
\end{center}
\end{figure}

Fig. \ref{masa} shows the lower limit to the NS mass as a function of $q$ 
obtained from $K_\mathrm{em} = 300$ \kms (lower solid line). Assuming 
$K_1\sim100$~\kms (favored by the diagnostic diagram) we obtain 
$q\lesssim0.35$, in agreement with \citet{Hynes2006}. On the other hand, if we 
use the maximum K-correction ($\alpha=0^{\circ}$) and the upper limit 
$K_\mathrm{em} = 320$~\kms we obtain an upper limit to the NS mass (upper 
solid line in Fig. \ref{masa}). \citet{Meyer1982} suggest that, even in the 
absence of irradiation, accretion discs in LMXBs have $\alpha \gtrsim 
6^{\circ}$ , which is also consistent with the results by \citet{D'Avanzo2006} 
for Cen X-4 in quiescence. Applying the the K-correction with $\alpha 
\gtrsim 6^{\circ}$ and $K_\mathrm{em} = 320$~\kms we obtain a less 
conservative upper limit represented by the dotted-dashed line in 
Fig. \ref{masa}. \par
\citet{Pearson2006} favored $q=0.34$ y $M_{1}\approx1.35\sol$ by fitting gas 
stream trajectories to the extended spot observed in the \hel{ii} Doppler map. 
This pair of values, which is marked by the dotted line in Fig. \ref{masa}, 
is not consistent with our results. Conversely, our results are consistent with those derived by \citet{Ozel2006}, which are represented by the grey region in Fig. \ref{masa}. Nevertheless, our values are also consistent 
with the presence of a canonical NS with $\sim 1.4\sol$ in EXO 0748-676. However, for this case to work we need $q\leq 0.24$

\subsection{The nature of the donor star}
The parameter constraint for EXO 0748-676 are consistent with a main sequence star and 
clearly a giant star cannot fit within the 3.8 h Roche lobe. 
\citet{Faulkner1972} showed that the mean density ($\bar{\rho}$) of a Roche 
lobe-filling companion star is determined solely by the binary period $P$:
\begin{center}
$\bar{\rho} \cong 113P_\mathrm{h}^{-2}$ g cm$^{-3}$
\end{center}
where $P_\mathrm{h}$ is the orbital period in hours.
For EXO 0748-676, with $P_\mathrm{h}=3.82$ hours, we obtain $M_2\sim0.42 \sol$ 
\citep{Cox2000}, which defines the right dashed line in Fig. \ref{masa} through 
$M_{1}=M_2/q$. Combining this relation with the lower, thick solid line we 
obtain $M_{1}\geq1.5\sol$ for the case of a main sequence donor.\\
Alternatively, the donor star could have evolved before the onset of mass  
transfer. For this to happen within a Hubble time, the binary must have 
experienced an episode of unstable mass transfer with $q>1$ in a thermal 
time scale. The final product of this scenario is a LMXB formed by a donor 
star with an evolved nucleus transfering material onto a NS. 
\citet{Schenker2002} have computed companion star masses versus the orbital 
period for the limiting case of a nearly all helium nuclei. 
Using their relation for the case of EXO 0748-676 we obtain $M_2\gtrsim 0.16\sol$ 
which yields the left dashed line in Fig. \ref{masa}. 
Altogether, the allowed region in the $M_{1}-q$ space of parameters yields 
$1\sol \leq M_{1} \leq 2.4\sol$ and $0.11\leq q \leq 0.28$ whereas a plausible 
$\alpha\geq 6^{\circ}$ implies $M_{1} \leq 2.0\sol$. It also interesting to consider a limiting case with
$q=0.28$, $\alpha=6^{\circ}$ and $\kem=320$~\kms together with $K_1 \sim 100$~\kms from the diagnostic diagram. For this case we obtain $q\geq0.22$ and therefore $M_1\geq1.3\sol$, favoring the main sequence scenario. However, we note that the unstable mass-transfer scenario was proposed to explain the observations in 
XTE J1118+480 \citep{Haswell2002} and Cyg X-2 \citep{Podsiadlowski2000}. 
More recently, \citet{Rodriguez-Gil2009} show that the mass of the 
companion star in the cataclysmic variable HS 0218+3229 is also in excellent 
agreement with the predictions of \citet{Schenker2002}. 
\section{Conclusions}
We present the first dynamical determination of the NS mass in EXO 0748-676. 
Assuming that the narrow components detected within the \he{ii} $\lambda4542$, 
$\lambda4686$ and $\lambda5412$ emission lines arise from the irradiated 
hemisphere of the donor we measure $K_\mathrm{em}=300\pm10$~\kms which results 
in $1\sol \leq M_{1} \leq 2.4\sol$ and $0.11\leq q \leq 0.28$. 
On the other hand, the mass of the donor star is constrained to  
$0.16\sol \leq M_{2} \leq 0.42\sol$ through unstable mass-transfer/main 
sequence models. For the case of a main sequence donor we find $M_{1} 
\geq 1.5\sol$.\par 
Our results are consistent with the \citet{Ozel2006} scenario and rule out 
other system parameters proposed by \citet{Pearson2006}. However, our
observations are also consistent with a canonical mass NS and more observations are
required to further constrain the mass of the NS in EXO 0748-676. Recently, 
EXO 0748+676 has been reported to reach quiescence at $R=21.9$ after 
$\simeq$22 years of activity \citep{Hynes2008}. This provides an excellent 
opportunity to measure the orbit of the companion star and determine the mass 
of the NS.
\vspace{1.5cm}

JC acknowledges support from the Spanish Ministry of Science and
Technology through the project AYA2007-66887.DS acknowledges a STFC Advanced Fellowship. Partially funded by the 
Spanish MEC under the Consolider-Ingenio 2010 Program grant CSD2006-00070: 
First Science with the GTC. The use of the spectral analysis 
software package MOLLY written by Tom Marsh is acknowledged.
\bibliographystyle{mn2e.bst}
\small

\bibliography{e0748_r}

\label{lastpage}
\end{document}